\documentclass[a4paper,11pt]{article}
\usepackage{pos}

\title{Seasonal variation of atmospheric muons}
 \ShortTitle{muon seaseonal}

\author*[a]{Thomas Gaisser}
\author[b]{Stef Verpoest}

\affiliation[a]{Bartol Research Institute, Dept. of Physics \& Astronomy\\
  University of Delaware, Newark, DE, USA}

\affiliation[b]{University of Gent, Dept. of Physics and Astronomy,\\
B-9000, Gent, Belgium}

\emailAdd{gaisser@udel.edu}
\emailAdd{stef.verpoest@ugent.be}

\abstract{Competition between decay and re-interaction of charged pions and kaons depends 
on the temperature/density profile of the upper atmosphere.  The amplitude and phase of
the variations depend on the minimum muon energy required to reach the detector and on
muon multiplicity in the detector.  Here we compare different methods for characterizing 
the muon production profile and the corresponding effective temperature.  A muon production 
profile based on a parameterization of simulations of muons as a function of primary energy 
is compared with approximate analytic solutions of the cascade equation integrated over 
primary energy.  In both cases, we compare two definitions of effective temperature.
We emphasize applications to compact underground detectors like MINOS and OPERA, while indicating how they relate to
extended detectors like IceCube.}

\FullConference{37$^{\rm{th}}$ International Cosmic Ray Conference (ICRC 2021)\\
		July 12th -- 23rd, 2021\\
		Online -- Berlin, Germany}

\begin{document}
\maketitle

\section{Introduction}
Two-body decays of $\pi^\pm$ and $K^\pm$ are the principal source of atmospheric
muons in the TeV energy range relevant for this paper, where the focus is on 
inclusive rates of muons from the steep spectrum of all cosmic rays.  Prompt muons
from decay of charm and three-body decays of kaons~\cite{Gaisser:2014pda} contribute 
significantly only at much higher energies, for example when primary energies in the 
PeV region and above can be selected by a surface array.  The relation of the muon rate to atmospheric temperature
evolves over a range determined by the critical energies for decay of
the parent mesons, $\epsilon_\pi = 115$~GeV, and $\epsilon_K = 857$~GeV.  At the lowest
energies both pions and kaons are below the threshold for re-interaction in the
atmosphere, so the correlation of the muon rate with temperature is small.  
Re-interaction becomes significant first for pions and only at higher energy for kaons.
Full correlation with temperature is reached for $E_\mu >> 1$~TeV.

The relation between measured muon rate and atmospheric temperature is conventionally
quantified by a correlation coefficient, $\alpha_T$,
\begin{equation}
\frac{\delta R}{R_{av}} = \alpha_T\frac{\delta T}{T_{av}},
\label{eq:alpha_T}
\end{equation}
where $T=T_{\rm eff}$ and $T_{av}$ is its average over a year and
$R$ is the rate of muons.  
Effective temperature is a convolution of the muon production
spectrum as a function of slant depth in the atmosphere with
the corresponding temperature profile.

The paper is organized with an initial section on the muon production.
We compare approximate analytic solutions of the hadronic cascade
equations with a muon production profile characterized by parameters determined from simulation.
The next section deals with effective temperature and compares two
approaches for relating temperature to muon production.  Finally, we
discuss calculation of the correlation coefficient, its evolution
with energy and how it varies between the different approaches to
calculation of rates and effective temperature.

\section{Rates of muons}
The rate of muons of energy $E_\mu$ from a direction $\theta,\phi$ 
in a detector with effective area $A_{\rm eff}$ is given by
\begin{equation}
{\rm R}(\theta,\phi) = \int{\rm d}X\int_{E_{\mu,min}}{\rm d}E_\mu\, A_{\rm eff}(E_\mu,\theta,\phi)P(E_\mu,\theta,X),
\label{eq:rtheta}
\end{equation}
where  $P(E_\mu,\theta,X)$
 is the production spectrum of muons differential in slant depth $X$.
 For a compact detector at a depth large compared to its vertical dimension,
 the effective area is the projected physical area from the direction $\theta,\phi$.
For simplicity, we consider detectors with a flat overburden, in which case  the physical 
area of the detector averaged over azimuth can be used and
\begin{eqnarray}
\label{eq:compact}
{\rm R}(\theta)& =&A_{\rm eff}(\theta) \int{\rm d}X\int_{E_{\mu,min}}{\rm d}E_\mu\,P(E_\mu,\theta,X) \\ \nonumber
&=&A_{\rm eff}(\theta)\int{\rm d}XP(>E_{\mu,min},\theta,X) \\ \nonumber
&=& A_{\rm eff}(\theta)\,I(E_{\mu,min},\theta),
\end{eqnarray}
where $I(E_{\mu,min},\theta)$ is the integral muon flux for a given zenith angle
and $E_{\mu,min}(\theta)$ is determined by the muon energy-loss formula and the
slant depth through the overburden for each zenith angle.
In both cases, the total rate is given by
\begin{equation}
{\rm Rate} = \sum_\theta {\rm R}(\theta).
\label{eq:rate}
\end{equation}

Here we use calculations for the MINOS Far Detector (FD) at Soudan~\cite{Adamson:2009zf} and the MINOS
ND at Fermilab~\cite{Adamson:2014xga} to compare two approaches to calculating the integral muon flux
and its dependence on atmospheric temperature in two different ranges of energy.
The standard approach is to use an analytic approximation for the 
integral flux of muons at slant depth $X$
\begin{equation}
P(>E_{\mu,min},\theta,X) = F(E_\mu)\frac{A_{\pi\mu}(X)}{\gamma+(\gamma+1)B_{\pi\mu}(X)E_\mu\cos\theta/\epsilon_\pi},
\label{eq:integralPmu}
\end{equation}
where $F(E_\mu)\equiv E_\mu\,N_0(E_\mu)$, and $N_0(E_\mu) = C\times E_\mu^{-(\gamma+1)}$ 
is the primary spectrum of nucleons per GeV m$^2$s sr evaluated at the energy of the muon. 
The integral spectral index is $\gamma\approx 1.7$, 
This form (plus the corresponding term for kaons) 
provides the production profile that can be inserted into Eq.~\ref{eq:compact} to get the integral
spectrum of muons.  The analytic form~\ref{eq:integralPmu} is based on a solution~\cite{Gaisser:2016uoy} to
the cascade equation for nucleons, pions and kaons in the atmosphere and produces
an inclusive muon flux that is not applicable to multiple muons.  The primary spectrum
is integrated assuming scaling for the production cross sections and a constant spectral
index and appears in Eq.~\ref{eq:integralPmu} evaluated at the energy of the muon.  The
production cross sections and two-body decays of the charged pions and kaons appear
as spectrum weighted moments for production and decay in the 
quantities $A$ and $B$ in Eq.~\ref{eq:integralPmu}:
\begin{equation}
A_{\pi\mu}(X)=\frac{Z_{N\pi}}{\lambda_N(\gamma+1)}\frac{1-r_\pi^{\gamma+1}}{1-r_\pi}e^{-X/\Lambda_N},
\label{eq:Api}
\end{equation}
and 
\begin{equation}
B_{\pi\mu}(X)=\frac{\gamma+2}{\gamma+1}\,\frac{1-r_\pi^{\gamma+1}}{1-r_\pi^{\gamma+2}}\,\frac{X}{\Lambda^*}
\,\frac{e^{-X/\Lambda_N}}{e^{-X/\Lambda_\pi}-e^{-X/\Lambda_N}},
\label{eq:Bpi}
\end{equation}
where $\Lambda_\pi^* = \Lambda_\pi\times\Lambda_N/(\Lambda_\pi-\Lambda_N)$ is a combination of the 
attenuation lengths for nucleons and pions.
The equations for the kaon channel have the same form, with the branching ratio
$0.635$ multiplying $A_{K\mu}(X)$.  For calculations we use the TeV values of 
spectrum-weighted moments and attenuations lenghts from Ref.~\cite{Gaisser:2016uoy}.

An alternate approach is to use a parameterization of Monte Carlo
simulations to calculate $P(>E_{\mu,min},\theta,X)$.  In this case, because the simulation
is following the production of muons along the trajectory of the primary cosmic ray, 
multiple muons are included.  In Ref.~\cite{Gaisser:2021cqh} the parameterization was applied
to the seasonal variation of multiple muon events as measured by MINOS~\cite{Adamson:2015qua}
and by the NOvA ND~\cite{Acero:2019lmp}.  Here we use it to calculate total rates of muons
integrated over the primary spectrum.  Because the total rates are
dominated by single muons, the comparison with the analytic approach is of interest.
For the primary spectrum in both cases we use the primary spectrum of nucleons from
the H3a model~\cite{Gaisser:2011cc,Gaisser:2013bla} of the spectrum and composition for the calculations shown below.

\begin{figure}
\centering
\includegraphics[width=0.7\textwidth]{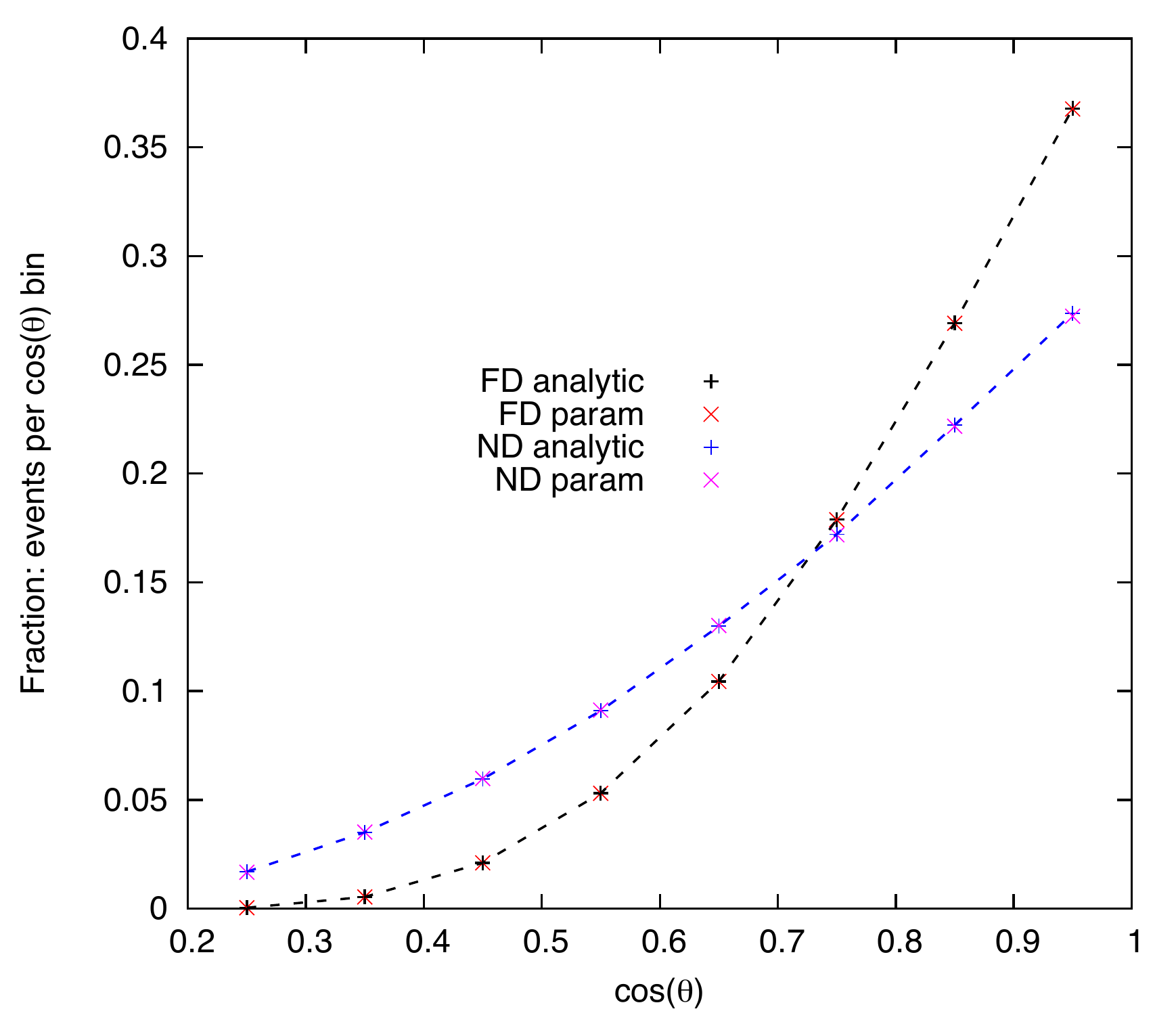}
\caption{Angular distribution the the MINOS near a far detectors.}
\label{fig:angular}
\end{figure}

Table~\ref{tab:Emin} shows the binning in zenith angle and the corresponding minimum 
muon energies used to calculate total rates (Eq.~\ref{eq:rate}).  The corresponding angular
distributions are shown in Fig.~\ref{fig:angular}.  The distribution is significantly flatter
for the shallow detector at Fermilab where the muons of lower energy are not so much influenced by 
radiative energy losses as for the deep detector at Soudan.
\begin{table}[htb]
   \caption{Minimum muon energies (GeV) for 8 bins of $\cos\theta$}
    \centering
    \begin{tabular}{r|c|c|c|c|c|c|c|c} \hline
      $\cos\theta$&0.95&0.85&0.75&0.65&0.55&0.45&0.35&0.25  \\ \hline
         MINOS FD&  730&850&1030&1320&1800&2730&5000&14000\\ \hline
         MINOS ND&50&56&64&74&89&111&147&217 \\ \hline
    \end{tabular}
     \label{tab:Emin}
\end{table}

\section{Effective temperature}
The effective temperature is a convolution of the atmospheric temperature profile with muon production
along a path defined by the zenith angle.  One possibility is to define it as
\begin{equation}
T_{\mathrm{eff}}(\theta)=\frac{\int{\rm d}X\,P(E_\mu,\theta,X)\,T(X)}{\int{\rm d}X\,P(E_\mu,\theta,X)},
    \label{eq:Teff}
\end{equation}
A simple derivation of a different definition of effective temperature starts by
taking the variance of the rate with respect to temperature.
\begin{eqnarray}
\Delta{\rm R}(\theta) &=& \int{\rm d}X\int{\rm d}E_\mu\, A_{\rm eff}(E_\mu,\theta) \\ \nonumber
&&\times\frac{{\rm d}P(E_\mu,\theta,X)}{{\rm d}T}\Delta T.
\label{eq:delta-rtheta}
\end{eqnarray}
Then define $\Delta T = T(X) - T_{\rm eff}$ and set $\Delta{\rm R} = 0$ to get
\begin{equation}
T_{\rm eff}(\theta) = \frac{\int{\rm d}X\int{\rm d}E_\mu\, A_{\rm eff}(E_\mu,\theta)
T(X)\frac{{\rm d}P(E_\mu,\theta,X)}{{\rm d}T}}{\int{\rm d}X\int{\rm d}E_\mu\, A_{\rm eff}(E_\mu,\theta)
\frac{{\rm d}P(E_\mu,\theta,X)}{{\rm d}T}}.
\label{eq:Teff2}
\end{equation}
This derivative definition of effective temperature originated with the first paper on 
seasonal variations of muons~\cite{Barrett:1952abc}, and a more recent implementation ~\cite{Grashorn:2009ey}
is used by MINOS and detectors such as OPERA~\cite{OPERA:2018jif} at LNGS.  Use of the simple
analytic approximation of Eq.~\ref{eq:integralPmu} leads to relatively simple forms listed in 
the next paragraph.  Application to the parameterization requires numerical differentiation of
$P(>E_\mu,\theta,Z)$.

The temperature dependence of the muon production spectrum is entirely contained
in the two critical energies,
\begin{equation}
\epsilon_i=\frac{m_ic^2}{c\tau_i}\frac{RT}{Mg}\,\,{\rm with}\,\frac{RT}{Mg}=29.62\,\frac{\rm m}{^\circ K}.
\label{eq:criticalE}
\end{equation}
Thus, for the differential form of the pion channel, for example,
\begin{equation}
T(X)\frac{dP(E_\mu,\theta,X)}{dT} = \frac{A_{\pi\mu}(X)B_{\pi\mu}(X)E_\mu\cos\theta/\epsilon_\pi(T)}
{[1 + B_{\pi\mu} E_\mu\cos\theta/\epsilon_\pi(T)]^2}.
\label{eq:dPdTanal}
\end{equation}
The corresponding integral form is
\begin{equation}
T_{\rm eff}(\theta) = \frac{\int{\rm d}X\
T(X)\frac{{\rm d}P(>E_\mu,\theta,X)}{{\rm d}T}}{\int{\rm d}X\,
\frac{{\rm d}P(>E_\mu,\theta,X)}{{\rm d}T}},
\label{eq:Teff1}
\end{equation}
with
\begin{equation}
T(X)\frac{dP(>E_\mu,\theta,X)}{dT} = \frac{A_{\pi\mu}(X)(\gamma+1)B_{\pi\mu}(X)E_\mu\cos\theta/\epsilon_\pi(T)}
{[\gamma + (\gamma+1)B_{\pi\mu} E_\mu\cos\theta/\epsilon_\pi(T)]^2}.
\label{eq:dPdTanal1}
\end{equation}

\begin{figure}[bht]
\centering
\includegraphics[width=0.45\textwidth]{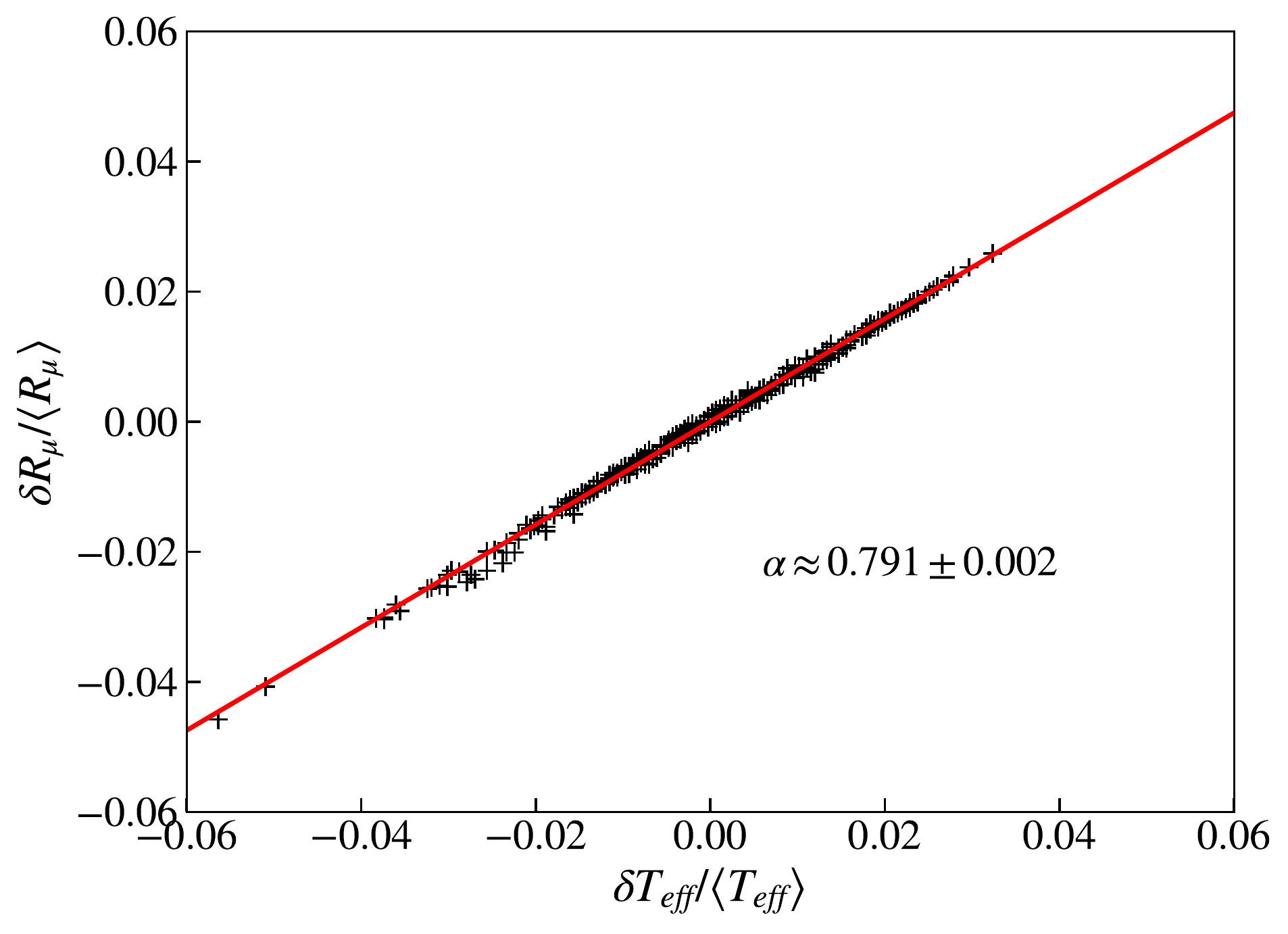}                                  \includegraphics[width=0.45\textwidth]{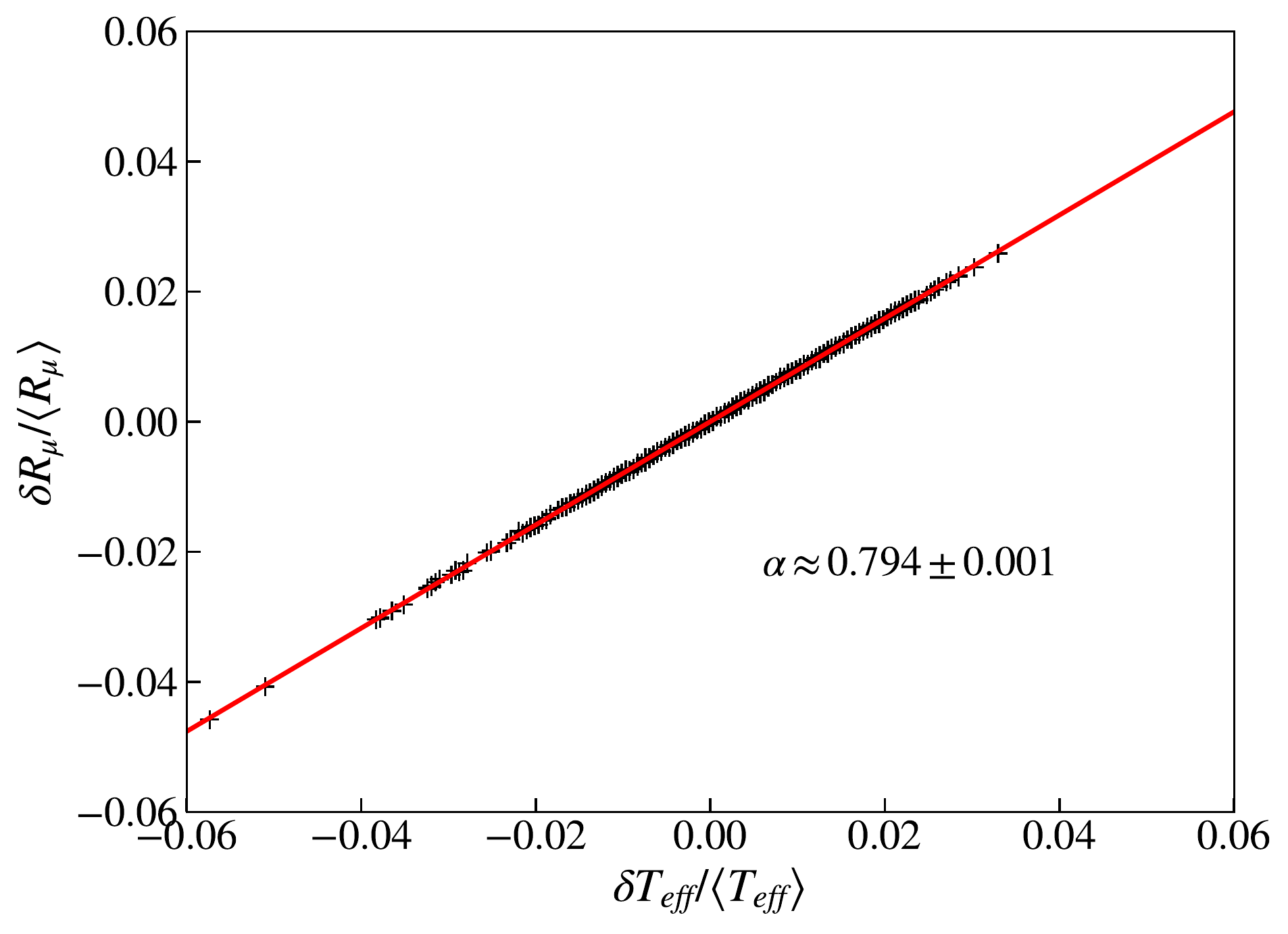}
\caption{Correlation with temperature for the MINOS FD calculated with the analytic formula; Left: T1 and Right: T2.}
\label{fig:analyticFD}
\end{figure}

\begin{figure}
\centering
\includegraphics[width=0.45\textwidth]{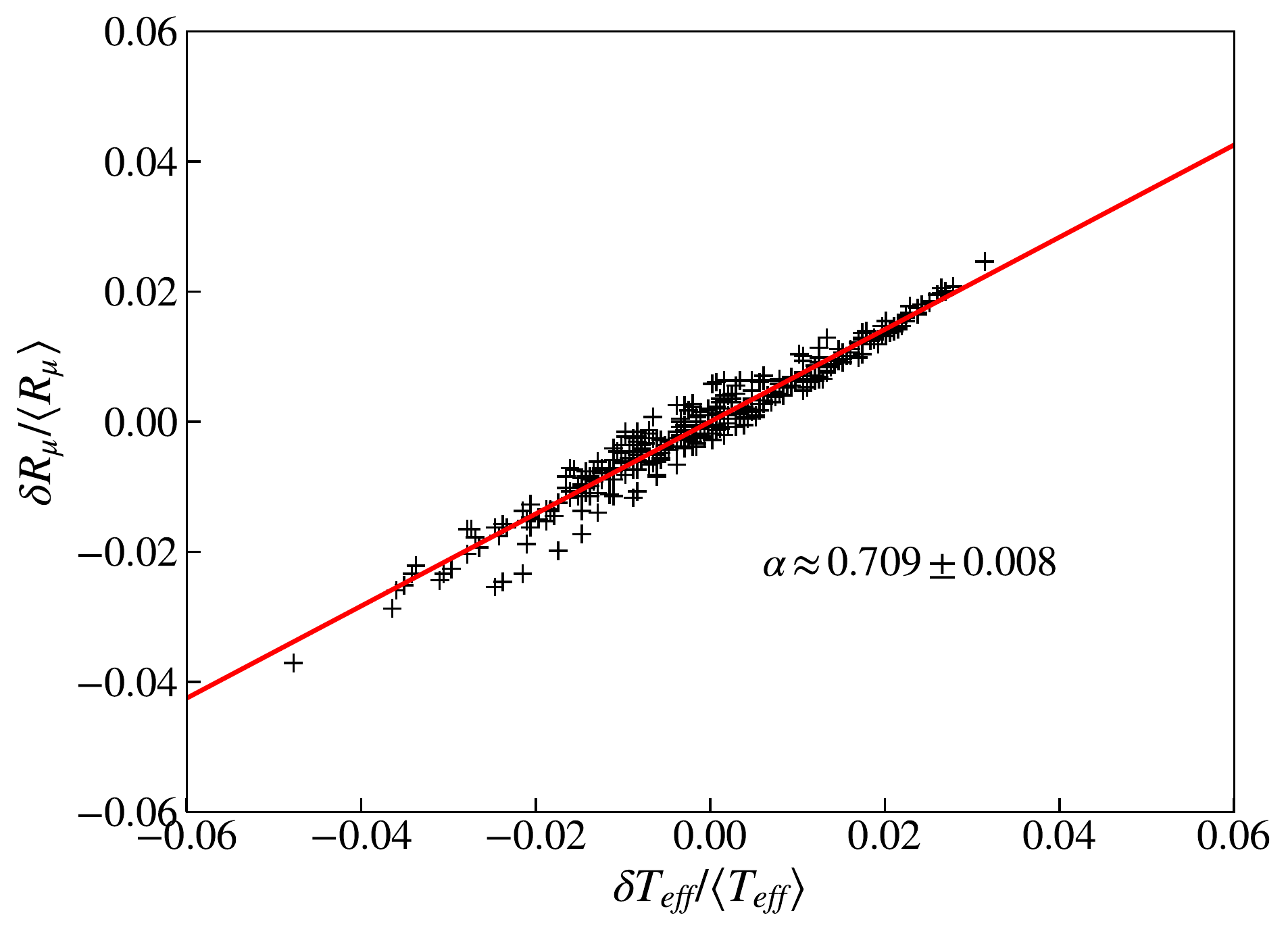}\includegraphics[width=0.45\textwidth]{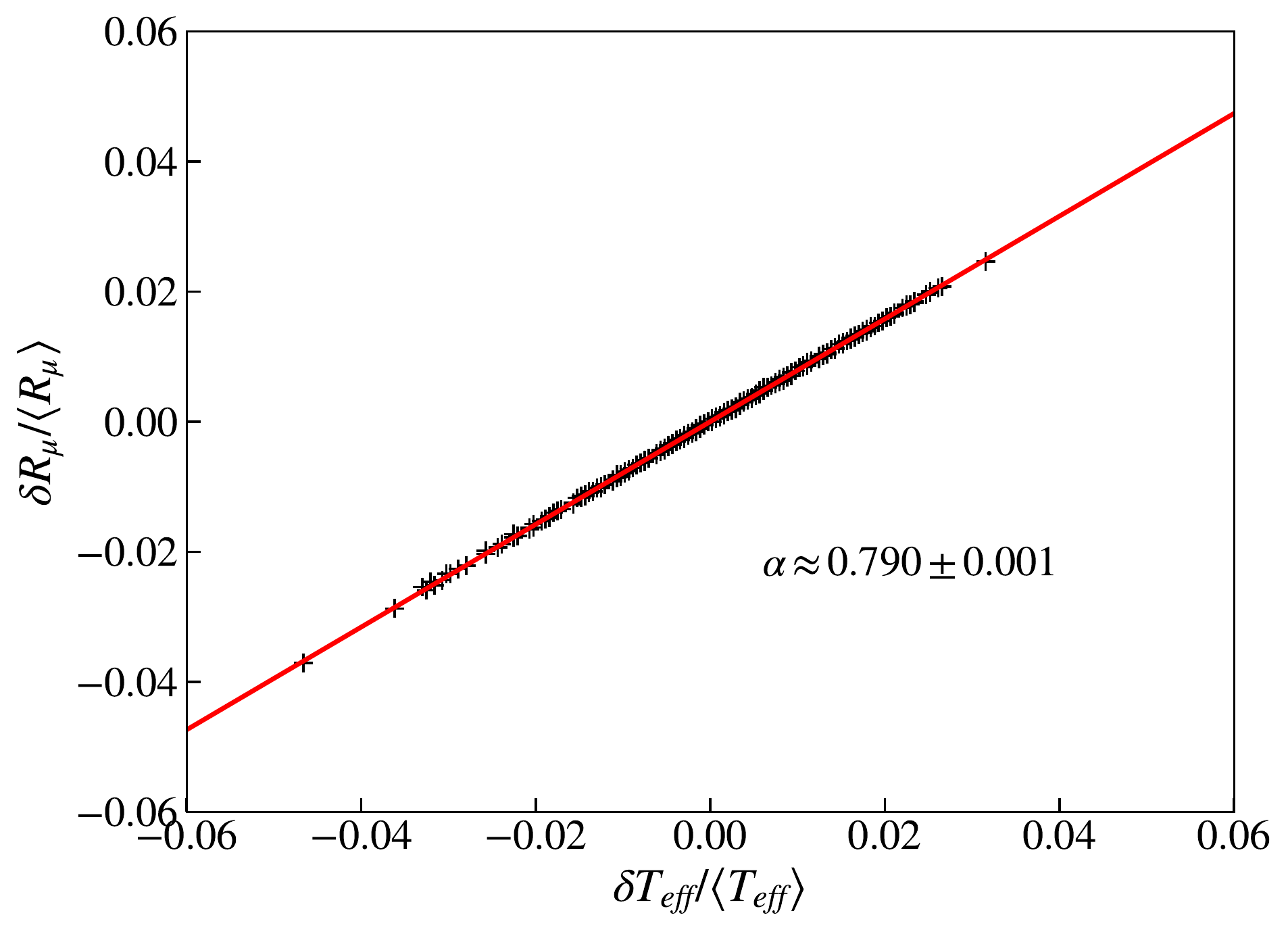}
\caption{Correlation coefficient for the MINOS FD calculated with the parameterization; Left: T1 and Right: T2.}
\label{fig:paramFD}
\end{figure}

\begin{figure}
\centering
\includegraphics[width=0.45\textwidth]{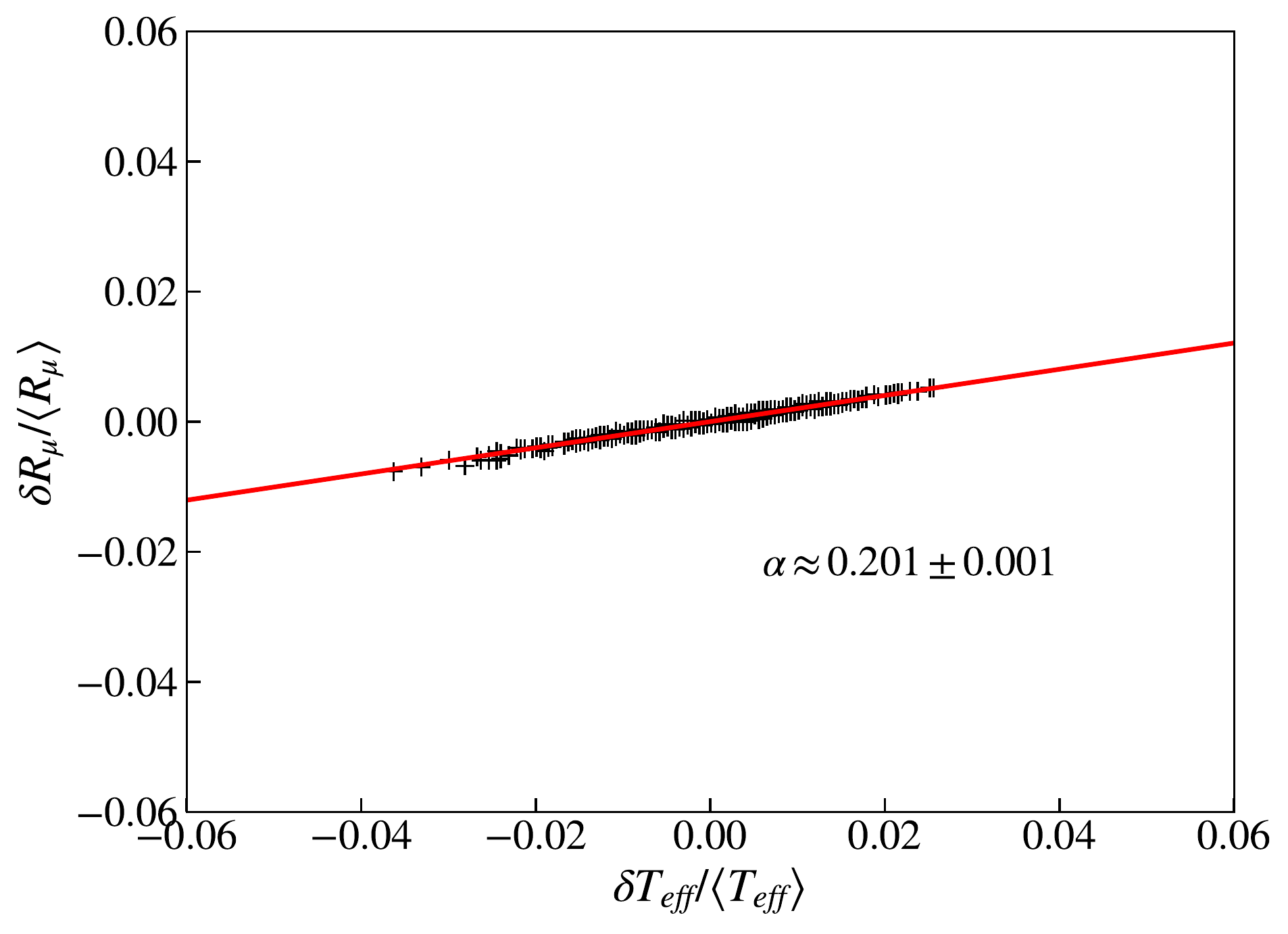}                                  \includegraphics[width=0.45\textwidth]{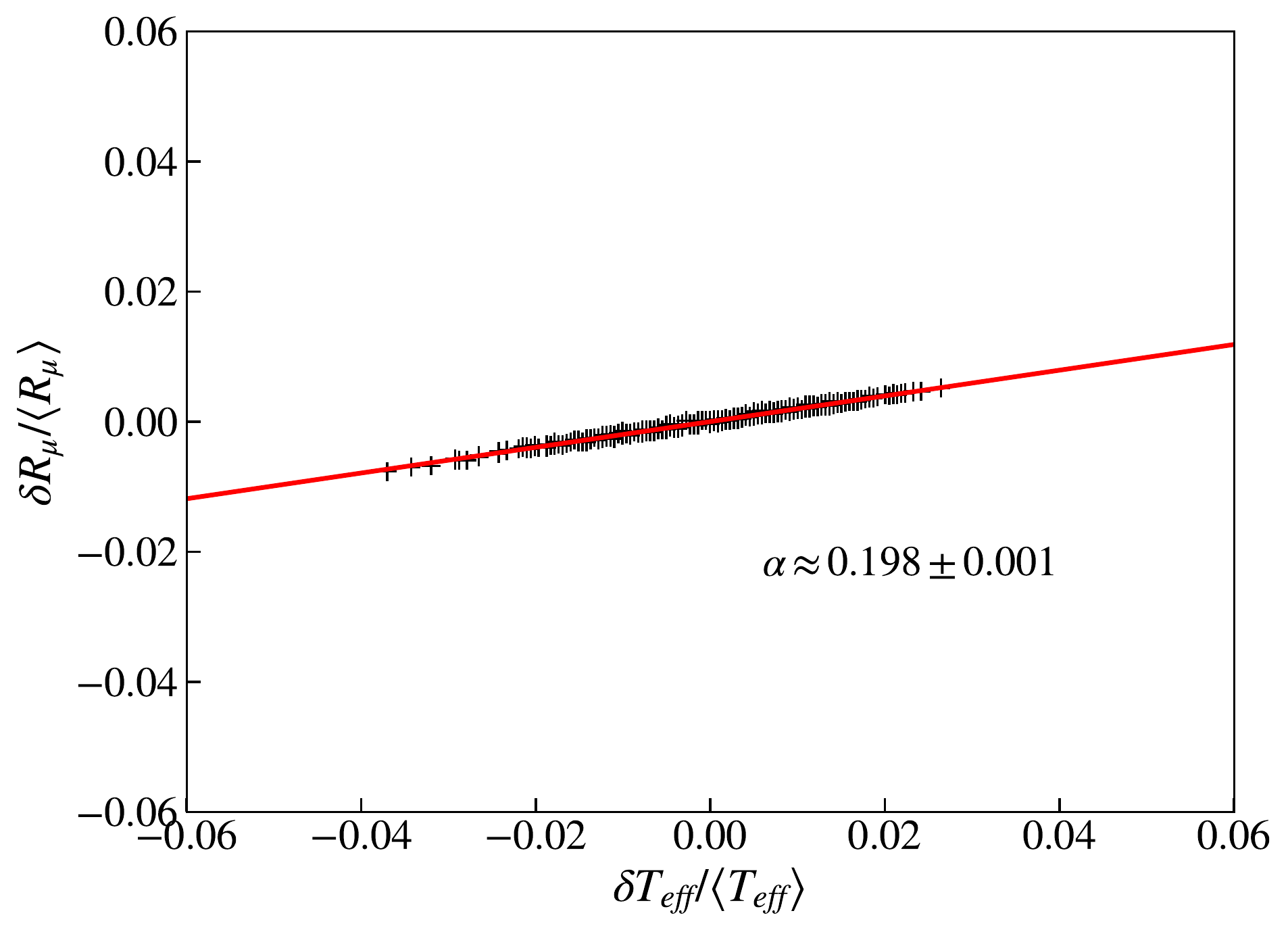}
\caption{Correlation with temperature for the MINOS ND calculated with the analytic formula; Left: T1 and Right: T2.}
\label{fig:analyticND}
\end{figure}

\begin{figure}
\centering
\includegraphics[width=0.45\textwidth]{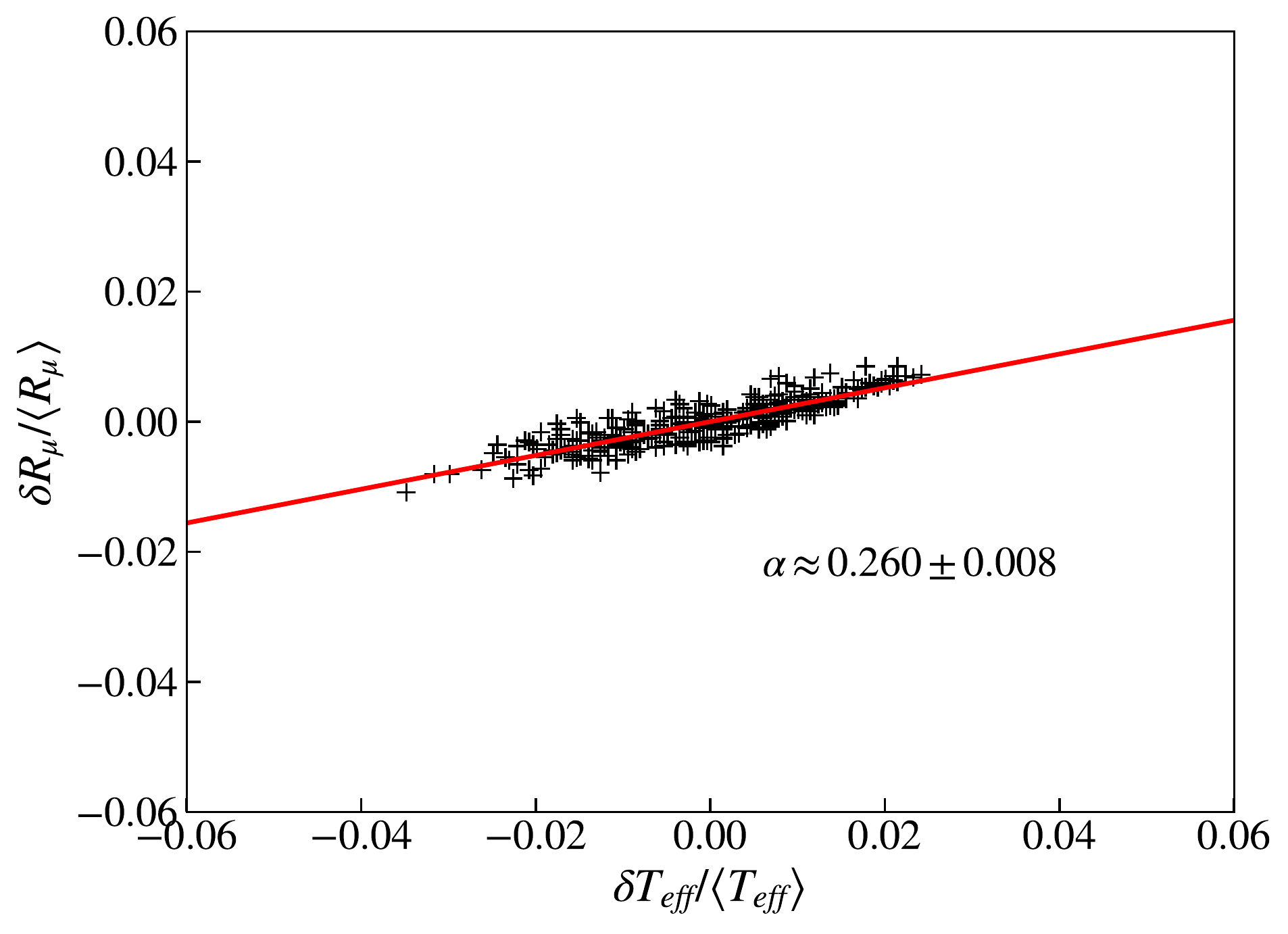}\includegraphics[width=0.45\textwidth]{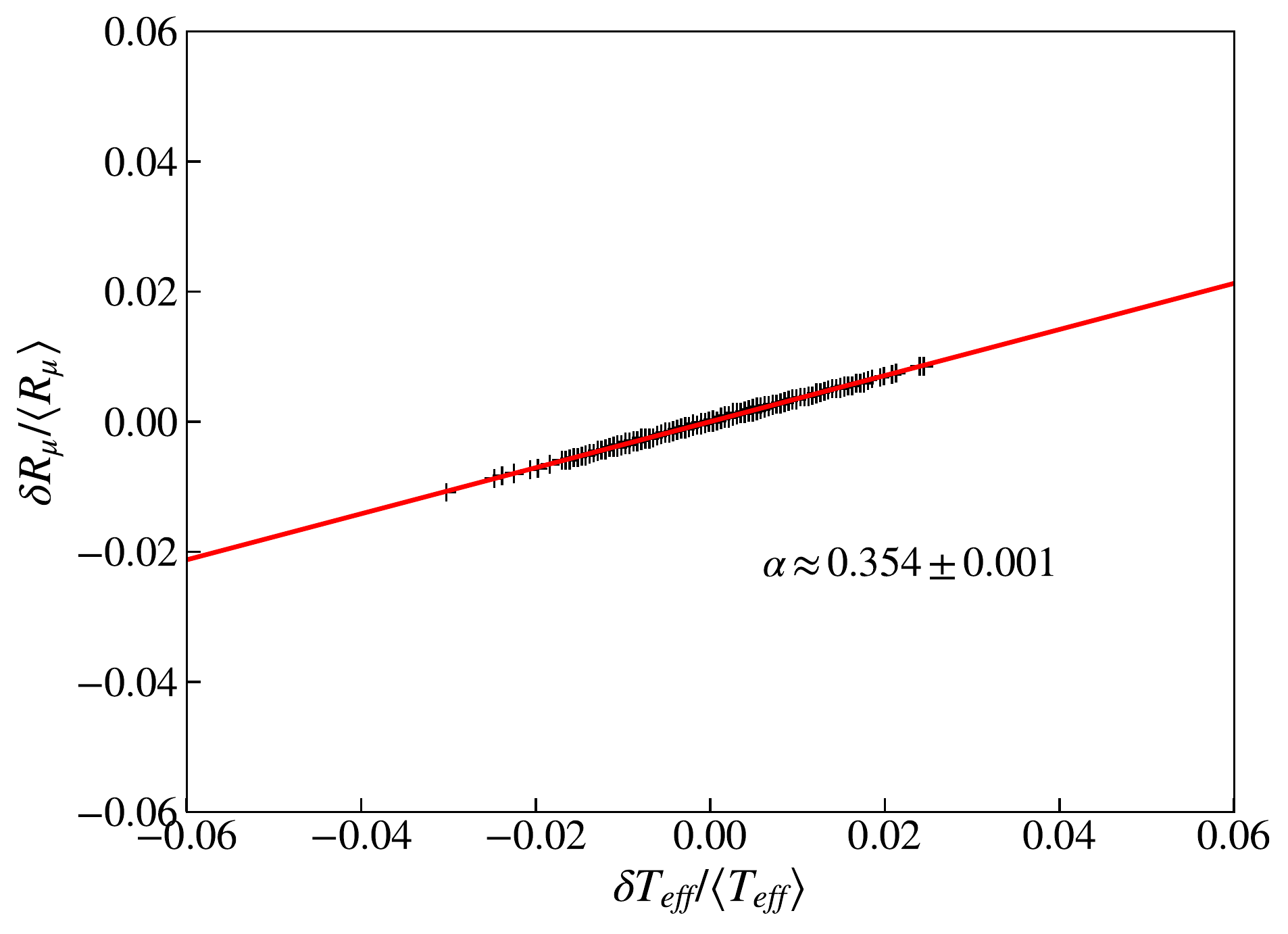}
\caption{Correlation coefficient for the MINOS ND calculated with the parameterization; Left: T1 and Right: T2.}
\label{fig:paramND}
\end{figure}

It is enlightening to apply the two definitions of effective temperature to
calculation of the correlation between rate and $T_{\rm eff}$.  Figure~\ref{fig:analyticFD}
shows the correlation from the analytic calculation for the MINOS FD.  The same comparison
using the parameterization is shown in Fig.~\ref{fig:paramFD}.  The corresponding
correlations for the MINOS ND are presented in Figs.~\ref{fig:analyticND} and~\ref{fig:paramND}.


\vspace{1cm}
\section{Correlation coefficient}

Slopes of straight-line fits to the data
shown in the correlation plots 
 give the respective coefficients, $\alpha_T$ from Eq.~\ref{eq:alpha_T}, for each set of assumptions.  In all cases, the derivative definition
 of effective temperature (Eq.~\ref{eq:Teff2}) gives a tighter correlation
 with the fit line.  This reduction of scatter is especially noticeable when
 the rates are calculated with the parameterization.
   For the MINOS ND the measured correlation coefficient is  $\alpha_{T} = 0.352\pm 0.046$~\cite{Adamson:2014xga},
 and for the FD $\alpha_{T} = 0.873$~\cite{Adamson:2009zf}.  It should be emphasized, however, that these are correlations with measured rates, whereas the figures here show correlations with calculated rates.  For the MINOS FD, all the calculations give coefficients somewhat below the measured value.
 It is interesting that for the shallow detector, the experimental value is closer to the result when the rates are calculated with
 the parameterization.

\section{Summary}

Understanding seasonal variations of muon rates in underground detectors
requires accounting for the temperature profile in the
integration of the muon production spectrum over slant depth in the atmosphere.
This paper compares two approaches to calculation of rates in compact 
underground detectors: 1) uses an analytic approximation to the muon production spectrum as a function of slant depth in the atmosphere and 2)
uses a formula based on simulations.  In both cases, the temperature-dependence is entirely contained in the critical energies
for pions and kaons.  Results are typically quantified as the correlation
between the muon rate and a single effective temperature evaluated for
each day (or other time span) for which the rates are determined, either
by calculation or by measurement.  We compared two definitions of $T_{\rm eff}$, one in which temperature is weighted by the muon production spectrum itself, and another in which it is weighted with the derivative  with respect to temperature of the
production spectrum.  The latter has been traditionally used in
analysis of results from compact underground detectors.  By definition
it minimizes deviation of calculated rates from the value expected for
a given effective temperature.

These different methods are illustrated in two different energy ranges 
by calculations for the MINOS ND ($E_\mu\sim 100$~GeV) and for 
the MINOS FD ($E_\mu\sim 1$)~TeV.  As expected, the correlation with 
effective temperature is significantly higher ($\alpha_T\sim 0.79$)
for the higher energy region than for lower energies ($0.2$ to $0.35$).
In the lower energy range, with $E_\mu\sim\epsilon_\pi$, the dominant pion component still has a high probability to decay, whereas in the 
higher energy range it is fully correlated with the temperature.

One aspect that remains to be examined is the relation between
the fractional contribution of the kaon channel to the calculated rates and the values of the corresponding correlation coefficients.  Other things
being equal, a larger kaon fraction should result in a smaller value of
the correlation coefficient.  The opposite is the case for the
lower energy calculation, where the kaon fraction is higher for the
parameterization, but the correlation coefficient is higher.

\vfill\eject
\bibliographystyle{ICRC}
\bibliography{G-V.bib}

\end{document}